# Three-dimensional Arbitrary Electromagnetic Fields and Temporal Propagation

**JORDAN M. ADAMS**[1,*] **AND DANIEL M. HELIGMAN**[2]

[1]*Riverside Research Institute, 2640 Hibiscus Way, Beavercreek, OH 45431*
*\*jadams@riversideresearch.org*

**Abstract:** We show that arbitrary 3D electromagnetic fields are transient solutions to Maxwell's equations and provide a simple equation to find how the field evolves over time. Multiple 3D fields can be realized at different times by superposing with an initial phase. Phase optimization algorithms allow for a phase-only modulated input signal. The necessary input wavepacket before a focus lens to create the 3D field can be found by finding the time-variation through a spatial plane. These results provide a way for designing arbitrary transient 3D waves and finding the wavepacket needed to input into a focusing lens.

## 1. Introduction

The amplitude and phase of light at a focal plane can be arbitrary crafted using spatial phase and amplitude modulation. This two-dimensional light shaping has benefited several applications such as particle manipulation [1-3] and biological stimulation and microscopy [4, 5]. The three-dimensional intensity distribution of light around a focus can also be arbitrarily shaped, however, with significant constraints. 3D focus shaping has numerous applications such as holographic displays [6-8], particle manipulation [9-11], rapid biological microscopic stimulation and imaging [12-14], and high throughput submicron additive manufacturing [15-17]. Holographic displays shapes light into surfaces to create visuals of objects while biological imaging and stimulation require sparse illumination. For these applications, 3D light shaping allows near arbitrary control of surfaces or sparsely distributed focal points over a large volume while maintaining high resolution. In comparison, multiphoton polymer and metal 3D printing have used 3D focus shaping to print complex objects without scanning a laser. In contrast to surface and spare illumination, additive manufacturing requires the focal field to have light shaped into solid objects. Crafting light into solid arbitrary shapes is only possible over a small volume as the large portion of light inside the objects will overlap with other regions as it propagates. One work suggested that 3D focal shaping could be improved by using polychromatic light and encoding different information into each frequency [18]. While this was correct that the target field is more accurately constructed in the 3D spatial volume, they failed to realize that encoding information in different frequencies also produces the target in the time variation through the spatial volume.

New experimental methods have shown it is possible to arbitrarily control the amplitude and phase of optical and near infrared wavepackets in space and time (2D+1) [18]. While this was a bulky setup, others have demonstrated compact 2D+1 spatiotemporal manipulation however for fixed and limited phase or amplitude patterns [21, 22]. It may not be long until a compact device is designed that can allow full 2D+1 phase and amplitude modulation of light in this frequency range. On the other hand, such 2D+1 modulation is already possible for acoustic waves [22] and lower frequency electromagnetic waves [23] using transducer or antenna arrays. With such capabilities, it will be valuable to understand the complex 3D+1 transient focal field that arises with tight focusing of complex wavepackets. It will be even more valuable to design an input wavepacket to a lens system based on an arbitrary 3D+1 transient target focal field.

In this paper, we show that arbitrary 3D fields are transient solutions to Maxwell's equations and can be propagated in time with a simple equation similar to the angular spectrum method (ASM) [24]. The optical system parameters like numerical aperture and the pulse bandwidth are shown to limit the resolution of the 3D transient focal field. Multiple 3D target fields can be realized at different times by superposing the targets with an initial phase. We provide simulation results showing time propagation exactly matches ASM. Finally, we use a phase retrieval algorithm to create a 3D transient focal field with phase-only modulation of a wavepacket. We believe this work can be used for crafting arbitrary transient forces for particle manipulation and particle acceleration. Additionally, this work can be used for shaping light to match complex 3D modes of molecules and crystals to excite exotic states of matter. For additive manufacturing, leveraging nonlinearity and accessory illumination it may be possible to use this method to obtain arbitrary time-averaged intensity profiles which can enable single exposure 3D printing over larger volumes for improved throughput. These results will also be valuable for controlling spatiotemporal field evolution for acoustic and lower-frequency electromagnetic waves.

## 2. Results

### 2.1 3D spatial transient fields and time propagation

The scalar wave-equation for the electric field, $E(x, y, z, t)$, in homogenous linear media is found from solving Maxwell's equations:

$$\nabla^2 E - \left(\frac{n}{c}\right)^2 \frac{\partial^2}{\partial t^2} E = 0 \tag{1}$$

where $n$ is the index of refraction and $c$ is the speed of light. Taking the 3D spatial Fourier transform gives an equation that describes the time variation of the spatial frequencies $\tilde{E}(k_x, k_y, k_z, t)$

$$-(2\pi)^2 k^2 \tilde{E} - \left(\frac{n}{c}\right)^2 \frac{\partial^2}{\partial t^2} \tilde{E} = 0 \tag{2}$$

with $k = \sqrt{k_x^2 + k_y^2 + k_z^2}$. Assuming the fields move one direction in time, the $\tilde{E}$ can easily be solved to be

$$\tilde{E}(k_x, k_y, k_z, t) = \tilde{E}_0 \exp\left(-i2\pi \frac{c}{n} t \sqrt{k_x^2 + k_y^2 + k_z^2}\right) \tag{3}$$

This equation describes how any arbitrary 3D spatial field at a given instance in time, $\tilde{E}_0 = \tilde{E}(k_x, k_y, k_z, t = 0)$, evolves over time. As time evolves, the 3D spatial frequencies simply obtain a phase a factor.

To illustrate this, we propagate a Benchy boat in time Fig. 1 (a)-(b), assuming the media is free-space with $n = 1$. The field value is set to one at the displayed locations and inside the boat (Fig. 1 (a) inset), giving a solid infill. We take the 3D FFT to obtain $\tilde{E}_o$ and propagate the field in time using eq. 3 which is shown Fig. 1 (b). Since there is no initial phase, the field is centered around zero spatial frequency (or DC) and the energy moves out spherically in all directions. Based on eq. 3, all spatial dimensions equally experience diffraction.

We give a more practical example in Fig. 1 (c)-(e) by adding a linear phase to the initial field to shift it to a commonly used Ti:Sapphire femtosecond laser center frequency of 375 THz or 800 nm center wavelength. We also constrain the temporal frequency bandwidth from 650 nm to 1100 nm and restrict the numerical aperture (NA) to 0.9 to simulate the available k-space bandwidth of tightly focused Ti:Sapphire femtosecond pulses. The field is set to zero outside of this k-space volume. The numerical aperture is sin $(\theta_k)$, where $\theta_k = \text{atan}\left(\frac{\sqrt{k_x^2+k_y^2}}{k_z}\right)$ is the k-space polar angle. The temporal frequency bandwidth creates a spherical shell as $\Delta\omega/c = \Delta k_r$ where $k_r = \sqrt{k_x^2 + k_y^2 + k_z^2}$ is the radius in k-space. Fig. 1 (d) shows that the constraints remove some high-frequency information from the field. With the shift in k-space to 375 THz, the field now propagates in the positive $z$ direction as time progresses (Fig. 1 (e)). As most of the energy is near at low spatial frequency in $k_y$ and $k_x$, the $x$ and $y$ profile slowly expands with propagation.

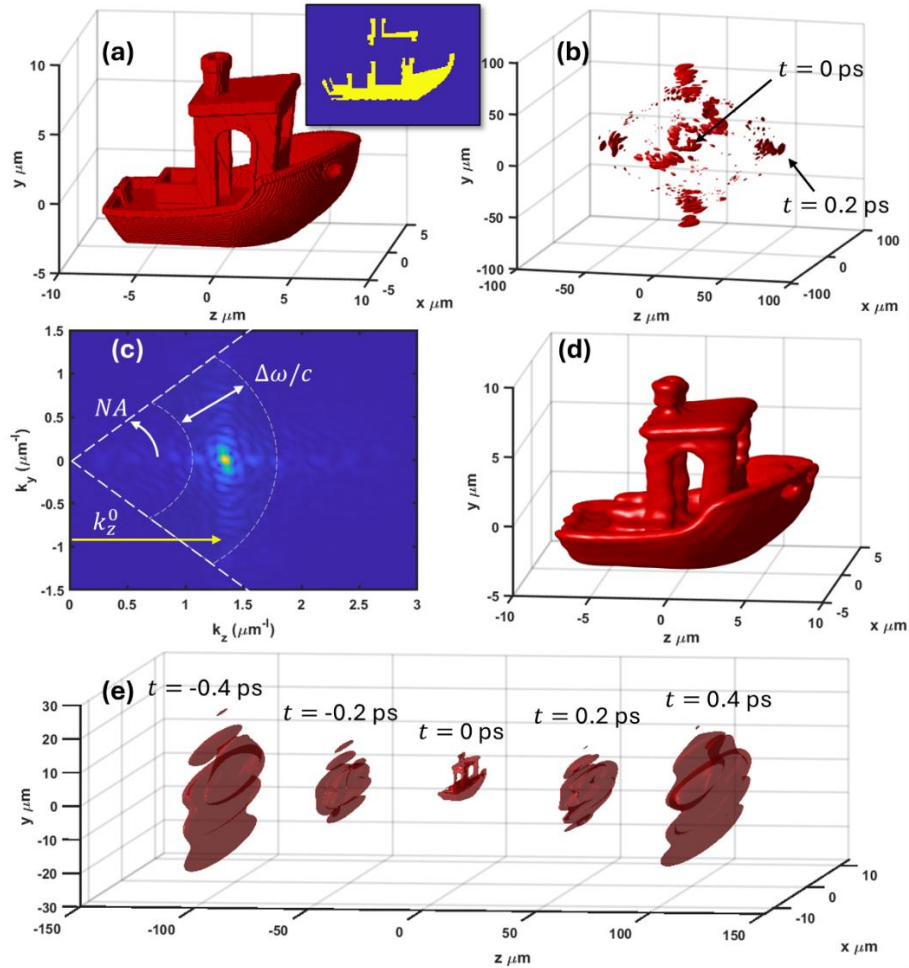

Fig. 1. Time propagation for unconstrained $k-$space bandwidth. (a) Iso-intensity plot of target 3D spatial field at $t = 0$ ps and inset cross-section showing the target is a solid volume. (b) Time propagation of a target field centered at zero center spatial frequency. (c) Shifting the spectrum to $\omega_0 = 375\ THz$ center frequency and applying angular limitations and temporal frequency bandwidth constraints. (d) 3D field at $t = 0$ ps for 0.9 NA and Ti:Sapphire femtosecond laser bandwidth limitations of 650 nm to 1100 nm. (e) Time propagation of 3D spatial field.

As the numerical aperture limits the angular span in k-space, it should control the angular spatial resolution. Similarly, the temporal frequency bandwidth should limit the radial spatial resolution. Fig. 2 (a)-(b) shows how field on the $y$-axis and on the $z$-axis is degraded as the temporal frequency bandwidth is lowered. Figure 2 (c)-(d) also shows that field on these axes is degraded as the numerical aperture is reduced. The temporal frequency bandwidth strongly affects the center axes as these are along the radial direction. However, despite being radially oriented the numerical aperture still degrades the field along the $y$ axis as it cuts off high-frequency energy in $k_y$, while the $z$ axis is less affected.

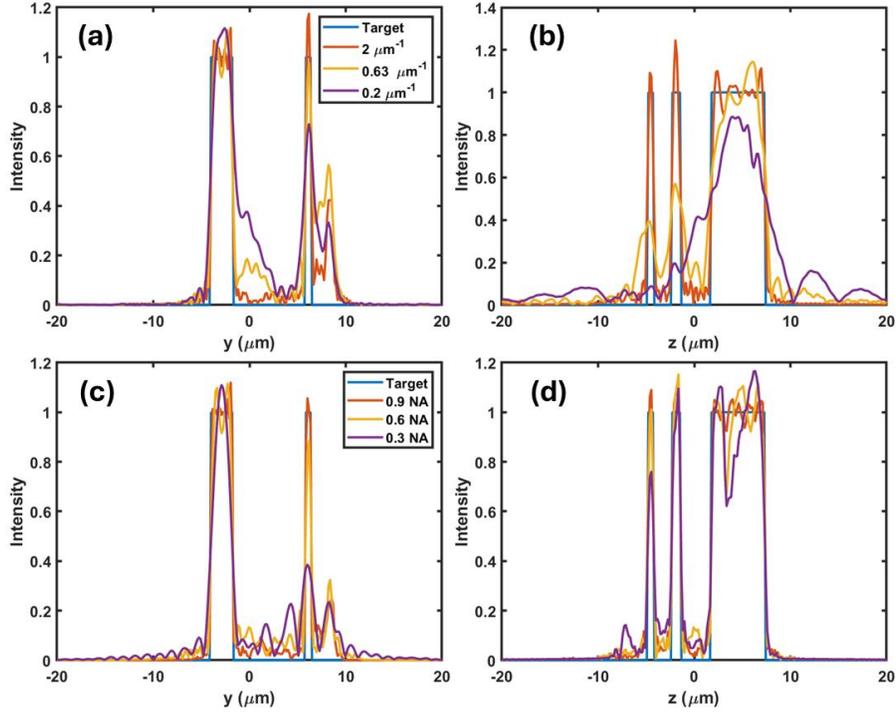

Fig. 2. Center-line fields for variable constraints on $k-$space bandwidth. Limiting temporal frequency bandwidth for 2 µm$^{-1}$, 0.65 µm$^{-1}$, and 0.2 µm$^{-1}$, with center frequency of 375 THz, for (a) $x = 0, z = 0$ and (b) $x = 0, y = 0$. Limiting spatial frequency bandwidth to 0.9 NA, 0.6 NA, and 0.3 NA for (c) $x = 0, z = 0$ and (d) $x = 0, y = 0$.

To verify this time propagation method accurately describes light propagation defined by Maxwell's equations, we compare results to the commonly used angular spectrum method (ASM) [24]. This is given by $E(k_x, k_y, z, \omega) = E(k_x, k_y, z = 0, \omega) \exp\left(i2\pi z \sqrt{\left(\frac{n\omega}{c}\right)^2 - k_x^2 - k_y^2}\right)$ which takes in any spatiotemporal field at a given $z$ plane and finds how the field evolves over $z$ by adding a phase in the spectral domain. Using time propagation, we first record the time-varying field through $z = 0$ of the propagating Bency boat without numerical or bandwidth constraints in Fig. 3 (a). From this time-varying field at $z = 0$, we use ASM to find the remaining field in $z$. The field derived from ASM at $t = 0 \ ps$ field is shown in Fig. 3 (b) which matches the initial field, proving time propagation accurately describes light propagation. Figure 3 (c)-(d) shows the same results when enforcing the 0.9 NA and 650 nm to 1100 nm bandwidth constraints. For these examples we used 0.4 fs time propagation steps. Reconstructing features at higher spatial frequencies with ASM will require increased time resolution in the $z$ plane.

The time varying field can also be used for finding the input wavepacket needed before a lens to produce the desired 3D transient field. The spatial profile of each frequency before a lens is the scaled inverse Fourier transform of the focus, where the scaling depends on both the lens focal length $f$ and the frequency $\omega$. As an alternative, the field could also be propagated using eq. 3 to find the time-variation through the $z = -f$ plane and the lens phase added to the spatial-spectral field to find the input wavepacket. For high numerical aperture focusing, apodization and electric field vector rotations from the lens must be accounted for to accurately find the input wavepacket [26].

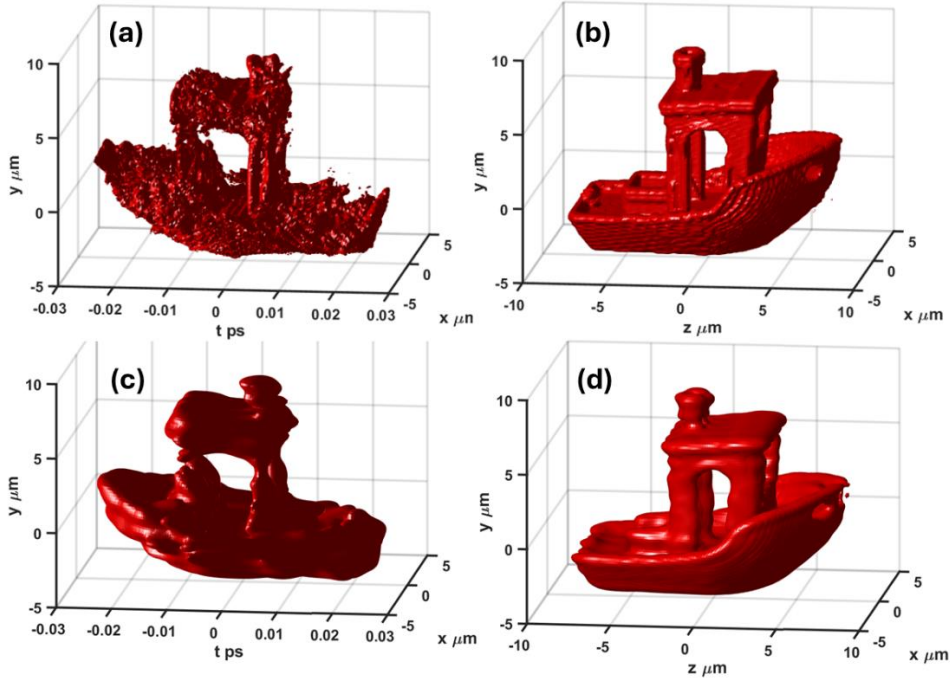

Fig. 3. (a) The time variation through the $z = 0$ plane for the field shifted to 375 THz with no numerical aperture or temporal frequency bandwidth constraints. (b) The spatial field at $t = 0$ reconstructed with ASM from $z = 0$ temporal field (0.4 fs time steps). (c)-(d) The same results for 0.9 NA and 650 nm to 1100 nm bandwidth limitations.

## 2.2 Superposition of multiple 3D fields

Multiple 3D target fields can be realized at different instances in time, if the fields do not overlap. For example, two targets $\tilde{E}_o^1$ and $\tilde{E}_o^2$ can be added together with phases based on their desired times of coming into focus at $t_1$ and $t_2$,

$$\tilde{E}_0 = \tilde{E}_0^1 \exp\left(i2\pi c t_1 \sqrt{k_x^2 + k_y^2 + k_z^2}\right) + \tilde{E}_0^2 \exp\left(i2\pi c t_2 \sqrt{k_x^2 + k_y^2 + k_z^2}\right). \tag{4}$$

As time evolves according to eq 3, the phase on $\tilde{E}_o^1$ will be canceled with the time-propagation phase at $t = t_1$. This is like a lens phase being added to a beam and the phase canceling a propagation phase at the focus.

As an example, we generate two fields at two instances in time. The first field, set at $t = 0$ ps, is a rectangular prism envelope with the Riverside Research logo carved into the center. The second field is the Benchy, set at $t = 0.23$ ps. Fig. 4 (a) plots the full spatial field at $t = 0$ ps showing the out-of-focus Benchy trailing the Riverside Research logo. Fig. 4 (b) plots the field at $t = 0.23$ ps showing the Benchy is now in focus, while the Riverside Research logo is out-of-focus. Again, given most of the energy is at lower spatial frequencies in $k_x$ and $k_y$ for these examples, the $x$ and $y$ profiles do not expand rapidly.

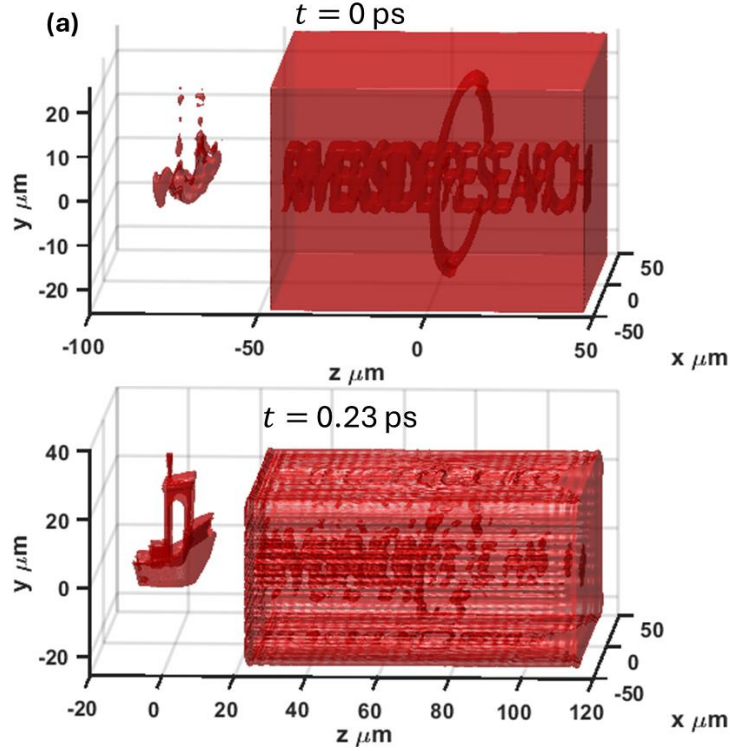

Fig. 4. The superposition of the Riverside Research logo and the Benchy boat, coming into focus at different locations constrained by 0.9 NA and 650 nm to 1100 nm bandwidth. (a) At $t = 0$ ps, the Riverside research logo comes into focus. (b) At $t = 0.23$ ps, the Benchy comes into focus.

### 3.3 Phase optimization for phase-only modulation

In the previous sections we simply shifted the target in k-space. A unique spatiospectral amplitude and phase input field would be needed before a lens to obtain the 3D transient field. However, full amplitude and phase modulation may be more difficult to realize for experiments and applications. For 3D targets that are only intensity profiles and have no requirments for the 3D phase, phase-only modulation can be used to give an approximate 3D field. For phase-only modulation, the amplitude in k-space can be contrained to a profile defined by the optical system and unmodulated wavepacket envelope, like a simple Guassian beam and approximate Gaussian spectral envelope. A necessary phase to add to the unmodulated wavepacket which will produce an approximated target field can be found using a Gerchburg-Saxton phase retrevial algorithm. Extending from planar versions used with ASM [24], the 3D algorithm is

$$\tilde{E}_0^i = FFT(E_0^i) \tag{5.1}$$

$$\tilde{E}_0^{i+1} = \exp(i\,\Gamma^i)\exp\left(-\frac{(k_r - k_0)^2}{w_r^2} - \frac{\theta_k^2}{w_\theta^2}\right), |\theta_k| < \sin^{-1}(NA) \tag{5.2}$$

$$E_0^{i+1} = iFFT(\tilde{E}_0^{i+1}) \tag{5.3}$$

$$E_0^{i+2} = E_0 \exp(i\gamma^{i+1}) \tag{5.4}$$

where $\Gamma^i = \text{angle}(\tilde{E}_0^i)$ and $\gamma^{i+1} = \text{angle}(E_0^{i+1})$. Additionaly, $w_r$ controls the radial bandwidth and $w_\theta$ controls the angular bandwidth in k-space. First the target field is taken to the spectral domain using the fast-fourir transform. Only the phase $\Gamma$ is kept and applied to the input wavepacket amplitude profile, which in this example is a Gaussian radially and angulalry with a NA cutoff (Fig. 5 (a)). This phase-only modulated field is inversed fast-fourier transformed to give the new spatial 3D field. The intensity profile is replaced with the target field, but the phase $\gamma$ is kept. This process can be interated several times to find an optimized k-space phase profile (Fig. 5 (b)) which results in the spatial field that closely matches the target as shown in Fig. 5 (c) and (d). Compared to the previous sections, the energy is now more evenly distrubted in $k_x$ and $k_y$ and the field expands rapidly as time advances (Fig. 5 (e)). This method guarantees the k-space field after a focusing lens is purely phase-modulated. Given the Fourier transform property of a lens, the input spatiospectral field before the lens will generally also be purely phase-modulated. While Ref. 19 attempted a polychromatic 3D phase retreival method to obtain a high-resolution static focal field, they failed to realize the light energy is always contrainted to surfaces in $k_x - k_y - k_z - \omega$ space and that encoding information in $\omega$ produces the target in space and time. In constrast, the 3D k-space of fields can be a volume in the time domain as discussed with these results.

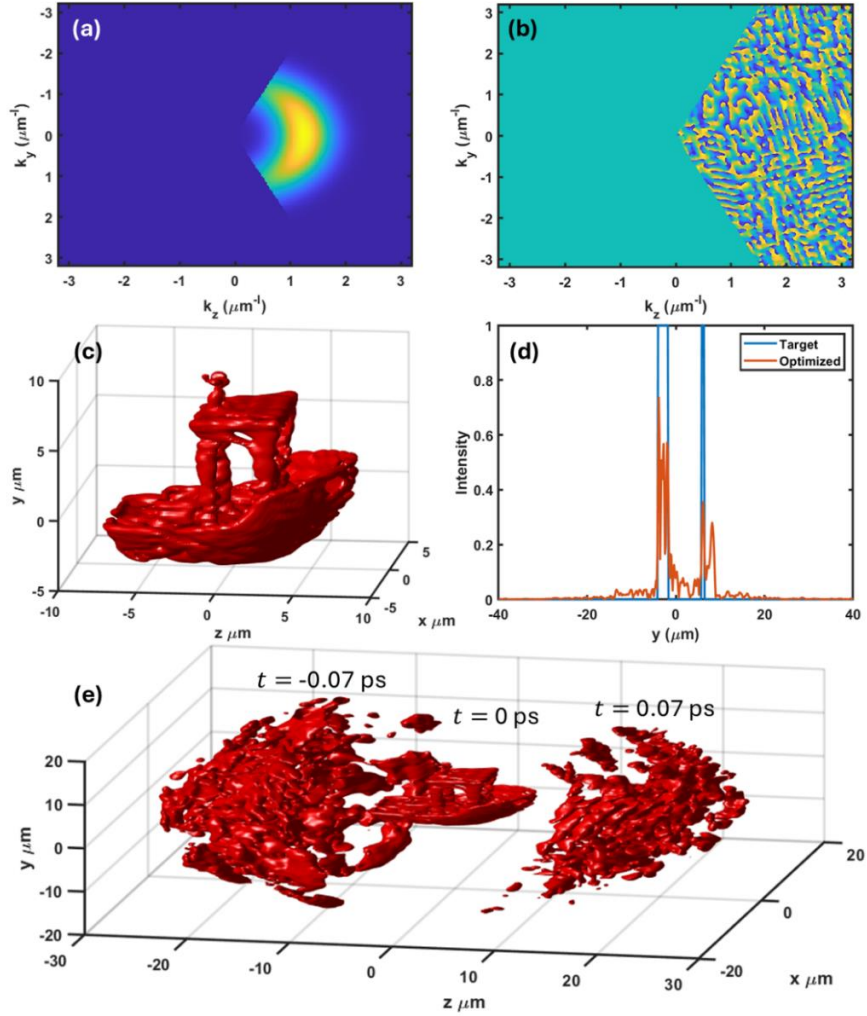

Fig. 5. (a) Radial and angular Gaussian envelope with 0.9 NA cut-off in k-space. (b) Optimized k-space phase to add to Gaussian envelope. (c) Resulting 3D field at $t = 0$ ps and (d) $x = 0$, $z = 0$ line profile. (e) Time propagation of the field.

## 3. Conclusion

In conclusion, we have shown that arbitrary 3D fields are transient solutions to the wave-equation and can be propagated in time by adding a temporal phase in the three-dimensional spatial frequency domain. Multiple fields can be realized at different times using superposition. Phase optimization algorithms allow for phase-only modulation in k-space. These results will be valuable for designing transient 3D fields and finding the necessary input wavepacket into a lens. This will benefit acoustic and lower-frequency electromagnetic wave applications which use transducer and antenna arrays to produce spatiotemporal fields, and will also benefit visible and NIR applications as 2D+1 phase and amplitude modulation techniques advance. These results could be used for crafting three-dimensional transient forces for particle manipulation and acceleration or exciting novel 3D spatial modes of matter.

**Funding.** Riverside Research

**Disclosures.** The authors declare no conflicts of interest.

**Data availability.** Data underlying the results may be obtained from the authors upon reasonable request.